\documentclass[preprint,nofootinbib,tightenlines,amsfonts,amsmath,amssymb]{revtex4}

\usepackage{bm}
\usepackage{color}
\usepackage{graphicx}
\usepackage{slashed}

\begin{document}
\baselineskip=15pt \parskip=4pt

\hspace{\fill} NCTS-PH/1709

\vspace*{3em}

\title{Exploring Spin-3/2 Dark Matter with Effective Higgs Couplings}

\author{Chia-Feng Chang$^{1}$}
\author{Xiao-Gang He$^{2,1,3}$}
\author{Jusak Tandean$^{1,3}$}

\affiliation{$^1$Department of Physics and Center for Theoretical Sciences, National Taiwan
University,\\ No.\,\,1, Sec.\,\,4, Roosevelt Rd., Taipei 106, Taiwan \smallskip \\
$^2$T.D. Lee Institute and School of Physics and Astronomy, Shanghai Jiao Tong University, \\
800 Dongchuan Rd., Minhang, Shanghai 200240, China \smallskip \\
$^3$Physics Division, National Center for Theoretical Sciences,
No.\,\,101, Sec.\,\,2, Kuang Fu Rd., Hsinchu 300, Taiwan \bigskip}


\begin{abstract}
We study an economical model of weakly-interacting massive particle dark matter (DM)
which has spin 3/2 and interacts with the 125-GeV Higgs boson via effective scalar
and pseudoscalar operators.
We apply constraints on the model from the relic density data, LHC measurements
of the Higgs boson, and direct and indirect searches for DM,
taking into account the effective nature of the DM-Higgs couplings.
We show that this DM is currently viable in most of the mass region from about 58 GeV to
2.3 TeV and will be probed more stringently by ongoing and upcoming experiments.
Nevertheless, the presence of the DM-Higgs pseudoscalar coupling could make parts
of the model parameter space elusive from future tests.
We find that important aspects of this scenario are quite similar to those of its more popular
spin-1/2 counterpart.
\end{abstract}

\maketitle

\section{Introduction\label{intro}}

Various astronomical and cosmological observations over the past several decades have led to
the wide acceptance that dark matter (DM) exists in our Universe, making up about 26\% of its
energy budget~\cite{pdg,Feng:2010gw}.
Despite the evidence, however, the identity of the basic constituents of the bulk of DM has
so far remained a mystery.
Since it cannot be accommodated by the standard model (SM) of particle physics, it is of much
interest to look into different possibilities beyond the SM which can offer good candidates
for DM.

Here we consider an economical scenario of DM which is of the popular weakly-interacting
massive particle (WIMP) type and has spin 3/2.
This kind of DM is still a viable alternative, although it has gained less attention than WIMP
candidates with spin 0, 1/2, or 1 in the literature ({\it e.g.}, Refs.~\cite{Baek:2014jga,
He:2016mls,Busoni:2014sya,deSimone:2014pda,Beniwal:2015sdl,Escudero:2016gzx,Arcadi:2017kky}).
Spin-3/2 WIMP DM could exist in a renormalizable new-physics model, such as the one proposed in
Ref.~\cite{Savvidy:2012qa}.\footnote{A well-known example for spin-3/2 DM is the gravitino in
supersymmetric theories (some possibilities of gravitino DM have recently been entertained in,
{\it e.g.}, \cite{gravitino}), but it is not regarded as a WIMP because it is extremely weakly
interacting and hence very hard to detect~\cite{pdg,Feng:2010gw}.\smallskip}
More model-independently, other analyses on spin-3/2 WIMP DM in recent years~\cite{Dutta:2015ega,
Kamenik:2011vy,Kamenik:2012hn,Yu:2011by,Chang:2017gla} have examined its potential interactions
with the SM sector via effective nonrenormalizable operators which involve other new states as
well~\cite{Dutta:2015ega} or only the DM and SM fields~\cite{Kamenik:2011vy,Kamenik:2012hn,
Yu:2011by,Chang:2017gla}.
In the following, we adopt the latter line of investigation assuming the absence of additional
nonstandard particles and focus specifically on the effective couplings of the DM to the standard
Higgs doublet.

This study is partly motivated by the null findings of the recent LUX~\cite{lux},
PandaX-II~\cite{pandax}, and XENON1T~\cite{x1t} direct detection experiments,\footnote{Overviews
on DM direct searches are available in~\cite{dmdd,Cushman:2013zza}.} which translate
into the most stringent upper-bounds to date on the cross section of spin-independent elastic
WIMP-nucleon scattering for WIMP masses between 4 GeV and 100 TeV.
These results imply major constraints on WIMP DM models, especially the simplest Higgs-portal
ones, which are also subject to restrictions from quests at the LHC~\cite{Aad:2015pla,atlas+cms}
for decays of the 125-GeV Higgs boson into final states which would signal the occurrence of
new physics.
In particular, as we demonstrated in Ref.\,\cite{Chang:2017gla}, spin-3/2 DM that links up
with the Higgs solely via an effective scalar operator is ruled out by the combination
of direct search and LHC data, except if the DM has a mass within a very small region
slightly below one half of the Higgs mass.
However, as also elaborated in Ref.\,\cite{Chang:2017gla}, if the model is enlarged somewhat
with another Higgs doublet, one could regain a good number of the eliminated masses.
In the present paper, we explore instead a different possibility in which we keep the minimal
particle content of the SM plus the spin-3/2 DM and suppose that the DM-Higgs interactions arise
not only from the effective scalar operator, but also from an effective pseudoscalar operator.
It turns out that this modification can provide the model with the freedom to evade
the preceding limitations over much of the mass region of concern.
Specifically, with the appropriate admixture of contributions from the scalar and
pseudoscalar couplings, the model can reproduce the observed DM relic abundance and
simultaneously yield DM-nucleon cross-sections which are low enough to allow for
the recovery of sizable parts of the excluded parameter space.
We will also look at a complementary probe of the model from indirect searches for DM.

In Sec.\,\ref{sec:sm+fdm}, we introduce the spin-3/2 WIMP DM, describe its couplings to
the Higgs boson, evaluate the model parameter values consistent with the relic density data,
and deal with the constraints on the DM from Higgs measurements and DM direct searches.
Since we assume that the interactions of the spin-3/2 DM with the Higgs are induced by
effective nonrenormalizable operators in the absence of other new particles, we also take
into account the restriction due to the limited extent of the reliability of
the effective-theory approximation.
Subsequently, we discuss specific examples of the viable parameter space of the model.
In Sec.\,\ref{comparison}, we make some comparison between this model and its spin-1/2 counterpart.
In addition, we briefly address how DM indirect detection experiments can offer extra
tests of these scenarios.
We give our conclusions in Sec.\,\ref{conclusion}.

\section{Higgs-portal spin-3/2 dark matter\label{sec:sm+fdm}}

The WIMP DM of interest is described by a Rarita-Schwinger field~\cite{rs},
denoted hereafter by a~Dirac four-spinor $\Psi_\nu$ with a vector index $\nu$.
For a free $\Psi_\nu$ with mass parameter $\mu_\Psi$, the Lagrangian can be expressed as
\begin{eqnarray} \label{L0}
{\cal L}_0^{} \,=\, \overline{\Psi_\kappa} L^{\kappa\nu}\Psi_\nu \,,
\end{eqnarray}
where~\cite{Moldauer:1956zz}\footnote{It is simple to obtain
\,$L^{\kappa\nu}=\big\{ i\slashed\partial-\mu_\Psi^{},\,[\gamma^\kappa,\gamma^\nu]\big\}/4
=\epsilon^{\kappa\rho\omega\nu\,}\gamma_{5\,}^{}\gamma_\rho^{}\partial_\omega^{}
- [\gamma^\kappa,\gamma^\nu]\mu_\Psi^{}/2$,\,
with \,$\epsilon_{0123}^{}=+1$.\,
These alternative formulas for $L^{\kappa\nu}$ are also employed in
the literature~\cite{Christensen:2013aua}.\medskip}
\begin{eqnarray} \label{Lkn}
L^{\kappa\nu} \,=\, -\big(i\slashed\partial-\mu_\Psi^{}\big)g^{\kappa\nu}
- \gamma^\kappa \big( i\slashed\partial+\mu_\Psi^{} \big) \gamma^\nu
+ i\gamma^\kappa\partial^\nu + i\gamma^\nu\partial^\kappa \,.
\end{eqnarray}
From ${\cal L}_0$, one finds the equation of motion \,$L^{\kappa\nu}\Psi_\nu=0$\, which
implies\footnote{More generally \cite{Moldauer:1956zz},
\,$L^{\kappa\nu}=-\big(i\slashed\partial-\mu_\Psi^{}\big)g^{\kappa\nu}
- i\texttt{\slshape A}_{\,}\gamma^\kappa\partial^\nu
- i\texttt{\slshape A}^*\gamma^\nu\partial^\kappa
- \gamma^\kappa \big( i\texttt{\slshape B}_{\,}\slashed\partial
+ \texttt{\slshape C}_{\,}\mu_\Psi^{} \big) \gamma^\nu$,\,
where the constant \texttt{\slshape A} ($\neq-1/2$) can be complex,
\,$\texttt{\slshape B}=1/2+{\rm Re}_{\,}\texttt{\slshape A}
+ 3|\texttt{\slshape A}|^2/2$,\,
and \,$\texttt{\slshape C}=1+3_{\,}{\rm Re}_{\,}\texttt{\slshape A}+3|\texttt{\slshape A}|^2$.\,
Thus the choice \,$\texttt{\slshape A}=-1$\, leads to\,\,Eq.\,(\ref{Lkn}).
It is straightforward to check that the relations in Eq.\,(\ref{eom}) are independent
of \texttt{\slshape A}.
Also, because of the condition \,$\gamma^\nu\Psi_\nu=0=\overline{\Psi_\nu}\gamma^\nu$,\,
quantities such as cross sections for exclusively on-shell $\Psi$ and $\bar\Psi$ do not
depend on \texttt{\slshape A}.}
\begin{eqnarray} \label{eom}
\gamma^\nu\Psi_\nu \,=\, 0 \,, ~~~ ~~~~ \partial^\nu\Psi_\nu \,=\, 0 \,, ~~~ ~~~~
\big(i\slashed\partial-\mu_\Psi^{}\big)\Psi_\nu \,=\, 0 \,.
\end{eqnarray}

For the DM processes to be evaluated in this section, the amplitudes are derived from
Feynman diagrams with the DM appearing only in external lines.
In such cases, the sum over the polarizations, $\varsigma$, of the DM particle
is~\cite{Christensen:2013aua,polsum}
\begin{eqnarray} \label{pols}
\raisebox{2pt}{\footnotesize$\displaystyle\sum_{\varsigma=-3/2}^{3/2}$}\,
u_{p,\varsigma}^\kappa\bar u_{p,\varsigma}^\nu
&=& (\slashed p + \mu_\Psi^{}) \biggl( -{\cal G}^{\kappa\nu}(p) +
\frac{\gamma_\rho\gamma_\omega}{3}\, {\cal G}^{\rho\kappa}(p)_{\,}{\cal G}^{\omega\nu}(p)
\biggr) \,,
\end{eqnarray}
where $p$ is its four-momentum and
\,${\cal G}^{\kappa\nu}(p)=g^{\kappa\nu}-p^\kappa p^\nu/\mu_\Psi^2$.\,
For the sum over the antiparticle's polarizations, the formula can be obtained from
Eq.\,(\ref{pols}) with the replacement \,$\mu_\Psi\to-\mu_\Psi$.\,

We assume that the DM candidate is stable due to an unbroken $Z_2$ symmetry under which
\,$\Psi_\nu\to-\Psi_\nu$\, and SM fields are not affected.
Furthermore, the lowest-order Higgs-portal interactions arise from dimension-five operators
containing the Higgs doublet \textsf{\slshape H} given by~\cite{Kamenik:2011vy}
\begin{eqnarray} \label{int}
{\cal L}_{\rm int}^{} \,=\, \Bigg( \frac{\overline{\Psi_\nu}\Psi^\nu}{\Lambda}
+ \frac{i\overline{\Psi_\nu}\gamma_5^{}\Psi^\nu}{\Lambda_5} \Bigg)
\textsf{\slshape H}^{\,\dagger\!}\textsf{\slshape H} \,,
\end{eqnarray}
where $\Lambda$ and $\Lambda_5$ are real constants which in general depend on
the couplings and masses characterizing the underlying heavy physics.
After electroweak symmetry breaking, the Lagrangian for $\Psi$ becomes
\begin{eqnarray} \label{Lint0}
{\cal L} \,=\, {\cal L}_0^{} + {\cal L}_{\rm int}^{} \,\supset\,
\overline{\Psi_\nu} \Bigg( \mu_\Psi^{} + \frac{v^2}{2\Lambda}
+ \frac{i\gamma_5^{}\,v^2}{2\Lambda_5} \Bigg) \Psi^\nu
+ \overline{\Psi_\nu} \bigg( \frac{1}{\Lambda} + \frac{i\gamma_5^{}}{\Lambda_5} \bigg)
\Psi^\nu \bigg(h v+\frac{h^2}{2}\bigg) \,,
\end{eqnarray}
where \,$\gamma^\nu\Psi_\nu=0$\, has been applied, $h$ refers to the physical Higgs boson,
and \,$v\simeq246$\,GeV\, is the vacuum expectation value (VEV) of \textsf{\slshape H}.
This indicates that the two effective operators induce corrections to the $\Psi$ mass.
A related effect is that, because of the $\Lambda_5$ contribution, $\Psi^\nu$ in
Eq.\,(\ref{Lint0}) is not yet a mass eigenstate.
Therefore, we need to perform the transformation
\,$\Psi_\nu\to e^{-i\gamma_5^{}\theta/2}\Psi_\nu$,\, so that we arrive at
\begin{eqnarray} \label{L0+Lint}
{\cal L} \,\supset\, m_\Psi^{}\overline{\Psi_\nu} \Psi^\nu
+ \overline{\Psi_\nu} \bigg( \frac{1}{\Lambda} + \frac{i\gamma_5^{}}{\Lambda_5} \bigg)
e^{-i\gamma_5^{}\theta}\Psi^\nu\, \bigg(h v+\frac{h^2}{2}\bigg) \,,
\end{eqnarray}
where now $\Psi^\nu$ is the mass eigenstate with mass
\begin{eqnarray}
m_\Psi^{} \,=\, \sqrt{\bigg(\mu_\Psi^{}+\frac{v^2}{2\Lambda}\bigg)^{\!2}
+ \frac{v^4}{4\Lambda_5^2}} ~,
\end{eqnarray}
which is connected to $\theta$ by
\begin{eqnarray}
\cos\theta \,=\, \frac{\mu_\Psi^{}}{m_\Psi^{}}+\frac{v^2}{2\Lambda m_\Psi^{}} \,, ~~~~ ~~~
\sin\theta \,=\, \frac{v^2}{2\Lambda_5^{}m_\Psi^{}} \,.
\end{eqnarray}
Consequently, $m_\Psi$ replaces $\mu_\Psi$ in Eqs.\,\,(\ref{eom}) and\,\,(\ref{pols}).
Moreover, the $\Psi$-$h$ interaction parts in Eq.\,(\ref{L0+Lint}) can be rewritten as
\begin{eqnarray} \label{Lint}
{\cal L} \,\supset\, \overline{\Psi_\nu} \big( \lambda_{\tt S}^{}
+ i\gamma_{5\,}^{}\lambda_{\tt P}^{} \big) \Psi^\nu\, \bigg(h+\frac{h^2}{2v}\bigg) \,,
\end{eqnarray}
with
\begin{eqnarray} \label{lambdaSP}
\lambda_{\tt S}^{} \,=\, \frac{v\cos\theta}{\Lambda}+\frac{v\sin\theta}{\Lambda_5} \,, ~~~ ~~~~
\lambda_{\tt P}^{} \,=\, \frac{v\cos\theta}{\Lambda_5} - \frac{v\sin\theta}{\Lambda} \,.
\end{eqnarray}

There are a couple of special cases worth mentioning.
If the $\Lambda_5$ term in Eq.\,(\ref{int}) is absent, corresponding to \,$\theta=0$,\,
then \,$\lambda_{\tt S}=v/\Lambda$\, and \,$\lambda_{\tt P}=0$,\, which is the case already
treated in Ref.\,\cite{Chang:2017gla} under the assumption of $CP$ conservation.
If there is no scalar coupling in Eq.\,(\ref{int}), due to \,$\Lambda\to\infty$,\, then
\begin{eqnarray*}
m_\Psi^{} \,=\, \sqrt{\mu_\Psi^2+\frac{v^4}{4\Lambda_5^2}} ~, ~~~~~
\lambda_{\tt S}^{} \,=\, \frac{v^3}{2\Lambda_{5\,}^2 m_\Psi^{}} \,, ~~~~~
\lambda_{\tt P}^{} \,=\, \frac{\mu_{\Psi\,}^{}v}{\Lambda_{5\,}^{}m_\Psi^{}} \,.
\end{eqnarray*}
One can see that with \,$1/\Lambda_5\neq0$,\, provided that \,$\Lambda\tan\theta\neq-\Lambda_5$,\,
there is always a nonvanishing contribution to $\lambda_{\tt S}$.
More generally, since we have taken $\mu_\Psi$, $\Lambda$, and $\Lambda_5$ to be free
parameters, $m_\Psi$ and $\lambda_{{\tt S},\tt P}$ are also free in what follows.

The couplings $\lambda_{{\tt S},\tt P}$ are responsible for the DM relic density, which
results from $\bar\Psi\Psi$ annihilation into SM particles, mainly
via the Higgs-mediated process \,$\bar\Psi\Psi\to h^*\to X_{\textsc{sm}}$.\,
If the center-of-mass energy $\sqrt s$ of the $\bar\Psi\Psi$ pair exceeds twice the Higgs
mass, $m_h^{}$, the channel \,$\bar\Psi\Psi\to hh$,\, due to contact and $h$-exchange
diagrams, has to be considered as well.\footnote{We have dropped contributions
to \,$\bar\Psi\Psi\to hh$\, from $t$- and $u$-channel $\Psi$-mediated diagrams because they
are of a~higher order in $\lambda_{{\tt S},\tt P}$ and of the same order as the potential
contributions of next-to-leading effective operators not included in Eq.\,(\ref{int}).}
Thus, the cross section $\sigma_{\rm ann}$ of DM annihilation is given by
\begin{eqnarray} \label{sigmaann}
\sigma_{\rm ann}^{} &=& \sigma\big(\bar{\Psi}\Psi\to h^*\to X_{\textsc{sm}}\big) \,+\,
\sigma\big(\bar{\Psi}\Psi\to hh\big) \,, \vphantom{|_{\int_\int^{}}}
\nonumber \\
\sigma\big(\bar{\Psi}\Psi\to h^*\to X_{\textsc{sm}}\big) &=&
\frac{\big(\eta_{\tt S}^{}\lambda_{\tt S}^2+\eta_{\tt P}^{}\lambda_{\tt P}^2\big)s^{5/2}\,
\raisebox{3pt}{\footnotesize$\displaystyle\sum_i$}\,\Gamma\big(\tilde h\to X_{i,\textsc{sm}}\big)}
{72_{\,}\beta_{\Psi\,}^{}m_\Psi^4\,\big[
\big(s-m_h^2\big)\raisebox{1pt}{$^2$}+\Gamma_h^2m_h^2 \big]} \,, ~~~~ ~~~
X_{i,\textsc{sm}} \,\neq\, hh \,,
\nonumber \\
\sigma\big(\bar{\Psi}\Psi\to hh\big) &=& \frac{\beta_h^{}
\big( \eta_{\tt S}^{}\lambda_{\tt S}^2+\eta_{\tt P}^{}\lambda_{\tt P}^2 \big) s^2}
{2304_{\,}\beta_{\Psi\,}^{}\pi_{\,}m_\Psi^{4\,}v^2}
\big({\cal R}_{hh}^2+{\cal I}_{hh}^2\big) \,,
\end{eqnarray}
where
\begin{eqnarray}
\beta_{\textsc x}^{} &=& \sqrt{1-\frac{4m_{\textsc x}^2}{s}} \,, ~~~~ ~~~
\eta_{\tt S}^{} \,=\, \frac{5\beta_\Psi^2-6\beta_\Psi^4+9\beta_\Psi^6}{8} \,, ~~~~ ~~~
\eta_{\tt P}^{} \,=\, \frac{9-6\beta_\Psi^2+5\beta_\Psi^4}{8} \,,
\nonumber \\
{\cal R}_{hh}^{} &=& 1 + \frac{3_{\,}m_{h\,}^2\big(s-m_h^2\big)}
{\big(s-m_h^2\big)\raisebox{1pt}{$^2$}+\Gamma_h^2m_h^2} \,, ~~~~ ~~~
{\cal I}_{hh}^{} \,=\, \frac{3_{\,}\Gamma_{h\,}^{}m_h^3}
{\big(s-m_h^2\big)\raisebox{1pt}{$^2$}+\Gamma_h^2m_h^2} \,.
\end{eqnarray}
Once $\lambda_{{\tt S},\tt P}$ have been extracted from the observed relic density, as
outlined in Ref.\,\cite{Chang:2017gla}, their values can be tested with various constraints.

If \,$m_{\Psi}^{}<m_h^{}/2$,\, the invisible channel \,$h\to\bar{\Psi}\Psi$\, is open.
We calculate its rate to be
\begin{eqnarray}
\Gamma\big(h\to\bar{\Psi}\Psi\big) \,=\, \frac{m_h^{}}{8\pi}~
\frac{\lambda_{\tt S\,}^2\big(1-4\texttt{\slshape R}_\Psi^2\big)
\big(1-6\texttt{\slshape R}_\Psi^2+18\texttt{\slshape R}_\Psi^4\big)
+ \lambda_{\tt P}^2
\big(1-2\texttt{\slshape R}_\Psi^2+10\texttt{\slshape R}_\Psi^4\big)}
{9\texttt{\slshape R}_\Psi^4} \sqrt{1-4\texttt{\slshape R}_\Psi^2} ~,
\end{eqnarray}
where \,${\texttt{\slshape R}}_f^{}=m_f^{}/m_h^{}$,\, in agreement with
Ref.\,\cite{Kamenik:2012hn}.
The LHC Higgs experiments can probe $\lambda_{{\tt S},\tt P}$ for
\,$m_{\Psi}^{}<m_h^{}/2$ via this decay mode.
According to the joint analysis by the ATLAS and CMS Collaborations of their
measurements~\cite{atlas+cms}, the branching fraction of Higgs decay into
channels beyond the SM is \,${\cal B}_{\textsc{bsm}}^{\rm exp}=0.00^{+0.16}$,\,
which can be interpreted as capping the branching fraction of~\,$h\to\bar\Psi\Psi$.\,
As a consequence, we may impose
\begin{eqnarray} \label{h2inv}
{\cal B}\big(h\to\bar\Psi\Psi\big) \,=\,
\frac{\Gamma\big(h\to\bar\Psi\Psi\big)}{\Gamma_h^{}} \,<\, 0.16 \,,
\end{eqnarray}
where \,$\Gamma_h^{}=\Gamma_h^{\textsc{sm}}+\Gamma\big(h\to\bar{\Psi}\Psi\big)$\, is
the Higgs' total width, which also enters the formulas in the last paragraph.
In numerical work, we set \,$m_h^{}=125.1$\,GeV,\, based on the current data~\cite{lhc:mh},
and correspondingly the SM width \,$\Gamma_h^{\textsc{sm}}=4.08$\,MeV \cite{lhctwiki}.

Another important test is available from direct detection experiments, which look for recoil
signals of nuclei due to the DM scattering off a nucleon, $N$, nonrelativistically at
momentum transfers that are small relative to the nucleon mass, $m_N$.
The relevant process is \,$\Psi N\to\Psi N$,\, which is mediated
by the Higgs in the $t$ channel.
Its cross section in the nonrelativistic limit is
\begin{eqnarray} \label{csel}
\sigma_{\rm el}^N \;=\; \frac{g_{NNh\,}^2 m_\Psi^2 m_N^2}{\pi_{\,}
\big(m_\Psi^{}+m_N^{}\big)\raisebox{0.7pt}{$^2$}m_h^4} \Bigg[ \lambda_{\tt S}^2 +
\frac{5_{\,}\lambda_{\tt P\,}^2m_{N\,}^2v_{\Psi,\rm lab}^2}
{18\big(m_\Psi^{}+m_N^{}\big)\raisebox{0.7pt}{$^2$}} \Bigg] \,,
\end{eqnarray}
where $g_{NNh}^{}$ parametrizes the effective Higgs-nucleon coupling defined by
\,${\cal L}_{NNh}=-g_{NNh\,}h\,\overline{N}N$\, and $v_{\Psi,\rm lab}$ denotes
the speed of the initial $\Psi$ in the laboratory frame.
Numerically, we adopt \,$g_{NNh}^{}=0.0011$\, following Ref.\,\cite{He:2016mls} and
\,$v_{\Psi,\rm lab}=300{\rm\,km/s}=10^{-3}$ \cite{pdg} relative to the speed of light.
The strongest restraints on $\sigma_{\rm el}^N$ to date for
\,$m_\Psi\mbox{\footnotesize$\,\gtrsim\,$}5$\,GeV\, are supplied by LUX~\cite{lux},
PandaX-II~\cite{pandax}, and XENON1T~\cite{x1t}.

In Eq.\,(\ref{csel}), the $v_{\Psi,\rm lab}^2$ factor clearly causes huge suppression in
the relative size of the $\lambda_{\tt P}$ and $\lambda_{\tt S}$ contributions to
$\sigma_{\rm el}^N$.
On the other hand, from Eq.\,(\ref{sigmaann}) we see that in the annihilation rate
\,$\sigma_{\rm ann}^{}v_{\rm rel}^{}$,\, where $v_{\rm rel}$ is the relative speed of
$\bar\Psi$ and $\Psi$ in their center-of-mass frame, the $\lambda_{\tt S}$ term is
suppressed by $v_{\rm rel}^2$ whereas the $\lambda_{\tt P}$ term is not,
as \,$\beta_\Psi\sim v_{\rm rel}/2$\, in the nonrelativistic limit.
This suggests that there may be admixtures of $\lambda_{\tt S}$ and $\lambda_{\tt P}$
contributions to $\sigma_{\rm ann}^{}$ and $\sigma_{\rm el}^N$ such that the various
pertinent requirements can be fulfilled.
Our evaluations below demonstrate that this is indeed the case.

Since we have $\lambda_{{\tt S},\tt P}$ as the free parameters besides $m_\Psi$,
it is convenient to express
\begin{eqnarray} \label{lambda}
\lambda_{\tt S}^{} &=& \lambda_{\Psi h\,}^{}\cos\xi \,, ~~~ ~~~~
\lambda_{\tt P}^{} \,=\, \lambda_{\Psi h\,}^{}\sin\xi \,, \vphantom{|_{\int_\int^{}}}
\nonumber \\
\lambda_{\Psi h}^{} &=& \sqrt{\lambda_{\tt S}^2+\lambda_{\tt P}^2} \,=\,
\sqrt{ \frac{v^2}{\Lambda^2} + \frac{v^2}{\Lambda_5^2} } \,.
\end{eqnarray}
Thus, for different values of \,$\lambda_{\tt P}/\lambda_{\tt S}=\tan\xi$\, we may explore
$(m_\Psi,\lambda_{\Psi h})$ regions complying with the aforesaid constraints.
However, as $\Lambda^{-1}$ and $\Lambda_5^{-1}$ belong to the effective operators in
Eq.\,(\ref{Lint}), we need to take into account the limited extent of validity of
the effective field theory (EFT) approximation.
To make a rough estimate for the EFT restraint on $\lambda_{\Psi h}$, we entertain
the possibility that each of the operators arises from a tree-level diagram mediated
by a heavy scalar $X$ having mass $m_X^{}$ and couplings to $\Psi$ and $h$ described by
${\cal L}_X^{}\supset-g_\Psi^{}\overline{\Psi_\nu}\Psi^\nu X-g_h^{}h^2X$
in the ultraviolet (UV) completion of the theory.
Moreover, inspired by the fermionic and scalar couplings of the SM, we suppose that
\,$g_\Psi^{}\sim\mu_\Psi^{}/v_X^{}$\, and \,$g_h^{}\sim\lambda_{hX}^{}v_X^{}$,\, where
$v_X^{}$ is the VEV of $X$ and $\lambda_{hX}$ is a\,\,constant, ignoring potential
modifications due to $h$-$X$ mixing.
The EFT will then remain reliable and perturbative if
\,$1/|\Lambda|\sim 2|\lambda_{hX}|\mu_\Psi/m_X^2<|\lambda_{hX}|/(2\mu_\Psi)<2\pi/\mu_\Psi$,\,
as the $s$-channel $\bar\Psi\Psi$ energy $\sqrt s$ satisfies
\,$m_X^2>s>4\mu_\Psi^2$\, and
\,$|\lambda_{hX}|<4\pi$\, for perturbativity.\footnote{The same bound,
\,$\Lambda>m_{\textsc{dm}}^{}/(2\pi)$,\, in the case of spin-1/2 DM was obtained in
\cite{Busoni:2014sya,Beniwal:2015sdl} employing similar arguments.}
Similarly,~$1/|\Lambda_5|<2\pi/\mu_\Psi$,\, although the heavy scalar may be different from $X$.
However, since the preceding bound on $\lambda_{hX}$ is its most relaxed, it is likely that the EFT
breaks down at significantly bigger $|\Lambda|$ and $|\Lambda_5|$, suggesting that it is
reasonable to demand instead \,$|\lambda_{hX}|<\cal O$(1).\,
Incorporating these into Eq.\,(\ref{lambda}) and additionally assuming
\,$\mu_\Psi^{}\sim m_\Psi^{}$,\, we can finally take \,$\lambda_{\Psi h}<v/m_\Psi^{}$.\,
In the \,$m_\Psi<m_h^{}/2$\, range, this restriction turns out to be much weaker than
that from Eq.\,(\ref{h2inv}) for the Higgs invisible decay, as will be seen shortly.

\begin{figure}[!b]
\includegraphics[width=81mm]{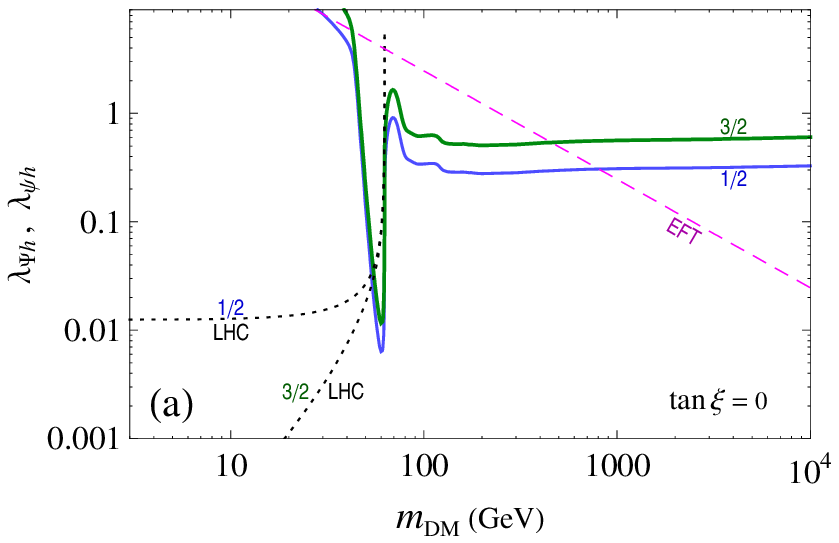} ~~
\includegraphics[width=83mm]{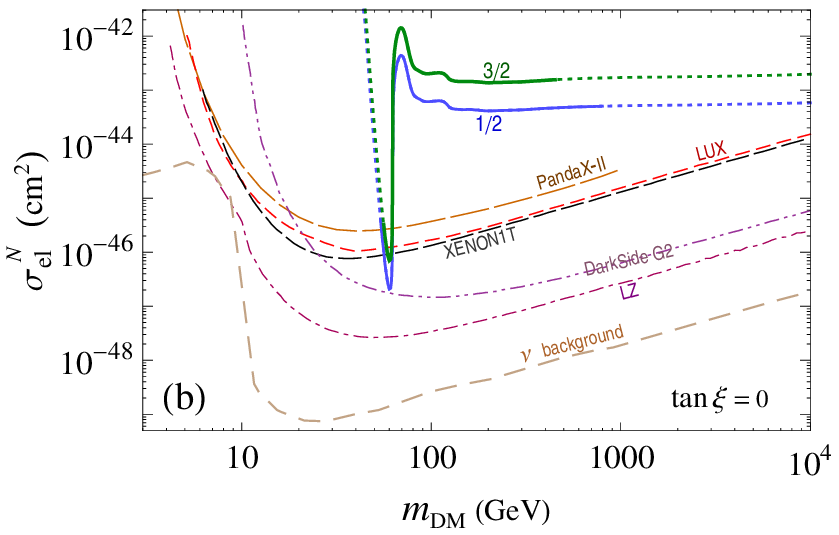}\vspace{-9pt}\\
\includegraphics[width=81mm]{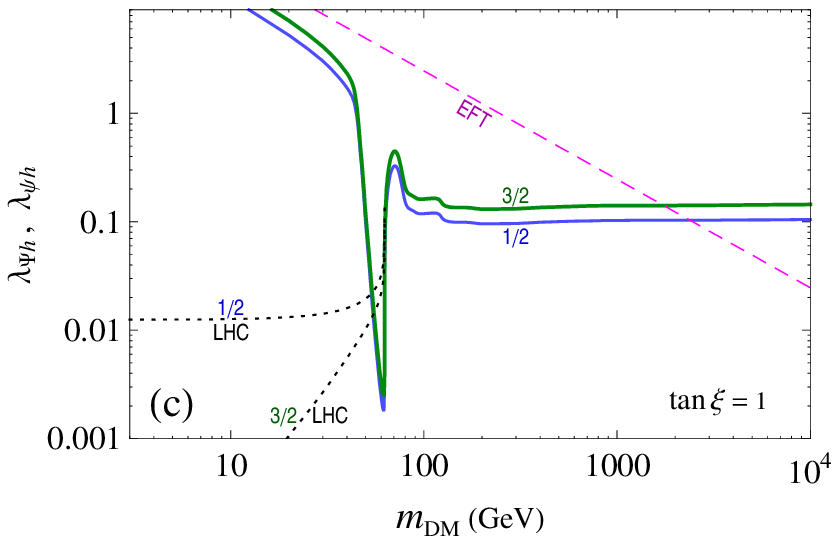} ~~
\includegraphics[width=83mm]{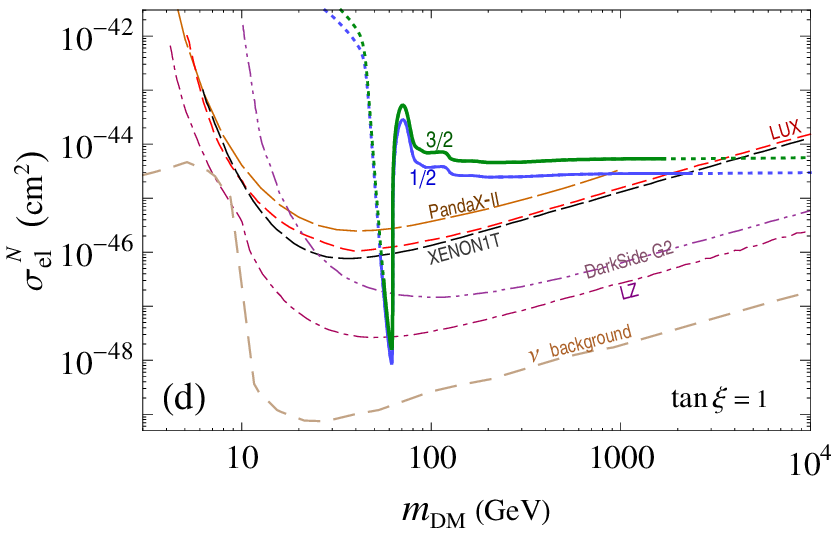}\vspace{-9pt}\\
\includegraphics[width=81mm]{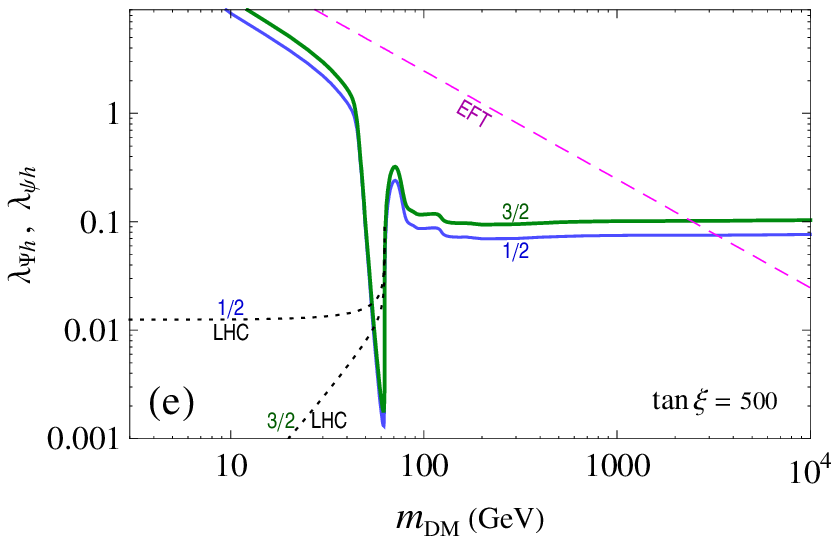} ~~
\includegraphics[width=83mm]{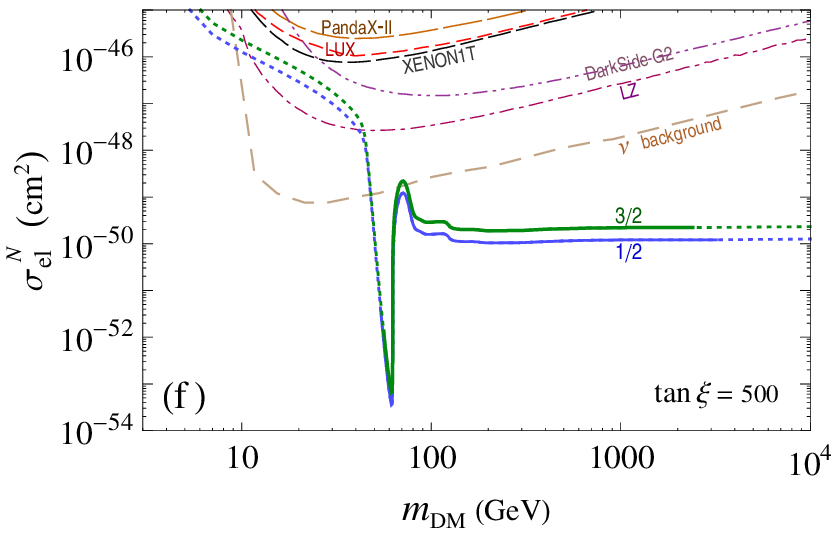}\vspace{-7pt}
\caption{\baselineskip=13pt%
Panels (a), (c), and (e): Values of the Higgs couplings $\lambda_{\Psi h}$ and $\lambda_{\psi h}$
of the spin-3/2 DM (green curves) and spin-1/2 DM (blue curves), respectively, versus the DM
mass which satisfy the relic density requirement for \,$\tan\xi=0,1,500$,\, compared to
the upper bounds inferred from LHC data on Higgs invisible decay (black dotted curves) and
from the limitation of the EFT approach (magenta dashed curves), as described in the text.
Panels (b), (d), and (f): The corresponding DM-nucleon cross-sections $\sigma_{\rm el}^{N}$
(green and blue curves), compared to the measured upper-limits from LUX~\cite{lux},
PandaX-II~\cite{pandax}, and XENON1T~\cite{x1t}, as well as the sensitivity
projections~\cite{Cushman:2013zza} of DarkSide\,\,G2~\cite{dsg2} and LZ~\cite{lz} and the WIMP
discovery lower-limit due to coherent neutrino scattering backgrounds~\cite{nubg}.
The dotted portions of the green and blue curves on the right are excluded by the LHC and EFT
restrictions in the left plots.} \label{sm+fdm-plots}
\end{figure}

Employing Eqs.\,\,(\ref{sigmaann}) and (\ref{csel}), we can determine the $\lambda_{\Psi h}$
values consistent with the observed relic density~\cite{planck} and predict the corresponding
$\Psi$-nucleon cross-section, $\sigma_{\rm el}^N$.
For a few representative choices of \,$\tan\xi=\lambda_{\tt P}/\lambda_{\tt S}$\, (namely
0, 1, and 500), we display
the results which are depicted by the green curves (labeled 3/2) in Fig.\,\,\ref{sm+fdm-plots}.
In the left plots, we also draw the upper limits on $\lambda_{\Psi h}$ inferred
from Eq.\,(\ref{h2inv}) based on the LHC Higgs data~\cite{atlas+cms}
(black dotted curves labeled 3/2 as well) and from \,$\lambda_{\Psi h}<v/m_\Psi^{}$\, for
the limited validity of the EFT description (magenta dashed curves).
The $\lambda_{\Psi h}$ values in the $m_\Psi$ ranges that meet the LHC and EFT requirements in
the left plots translate into the solid portions of the green curves for $\sigma_{\rm el}^N$
in the right plots.
In each of the right plots, the theoretical $\sigma_{\rm el}^N$ can be compared to the measured
upper-limits at 90\% confidence level (CL) from LUX~\cite{lux} (red dashed curve),
PandaX-II~\cite{pandax} (orange long-dashed curve), and XENON1T~\cite{x1t} (black medium-dashed
curve), as well as the sensitivity projections~\cite{Cushman:2013zza} of the future experiments
DarkSide\,\,G2~\cite{dsg2} (purple dash-dot-dotted curve) and LZ~\cite{lz} (maroon
dash-dotted curve) and the WIMP discovery lower-limit due to coherent
neutrino scattering backgrounds~\cite{nubg} (brown dashed curve).

It is evident from Figs.\,\,\ref{sm+fdm-plots}(a), \ref{sm+fdm-plots}(c), and \ref{sm+fdm-plots}(e)
that as \,$\lambda_{\tt P}/\lambda_{\tt S}=\tan\xi$\, exceeds unity the effect of $\lambda_{\tt S}$
on the annihilation rate quickly becomes negligible, in agreement with expectation based on
the $v_{\rm rel}$ suppression of the $\lambda_{\tt S}$ terms relative to the $\lambda_{\tt P}$
terms in the annihilation cross-section in Eq.\,(\ref{sigmaann}).
Accordingly, when $\lambda_{\tt P}/\lambda_{\tt S}$ grows large, the $\lambda_{\Psi h}$
values (green curves) become independent of this ratio.
More interestingly, the instances in Fig.\,\,\ref{sm+fdm-plots} for the spin-3/2 DM illustrate
that with \,$\lambda_{\tt P}\neq0$\, it is possible to recover at least some of the parameter
space excluded by the direct-search limits in the \,$\lambda_{\tt P}=0$\, case
[Fig.\,\,\ref{sm+fdm-plots}(b)] and perhaps even to escape future ones.
Especially for \,$m_\Psi>50$\,GeV,\, our numerical computations reveal that at present the strictest
bound from XENON1T is completely evaded if \,$\lambda_{\tt P}>25\,\lambda_{\tt S}$.\,
Moreover, in this $m_\Psi$ region, most of the predictions for $\sigma_{\rm el}^N$ are below
the neutrino-background floor if \,$\lambda_{\tt P}>500\,\lambda_{\tt S}$,\,
which is exhibited in Fig.\,\,\ref{sm+fdm-plots}(f).
However, our calculations further show that for \,$\lambda_{\tt P}>25\,\lambda_{\tt S}$\,
the LHC and EFT restrictions can be fulfilled only within the range
\,58\,\,GeV\,$\mbox{\footnotesize$\lesssim$}\,m_\Psi\,
\mbox{\footnotesize$\lesssim$}\;$2.3\,\,TeV,\,
as Fig.\,\,\ref{sm+fdm-plots} also indicates.

\begin{figure}[h]
\includegraphics[width=83mm]{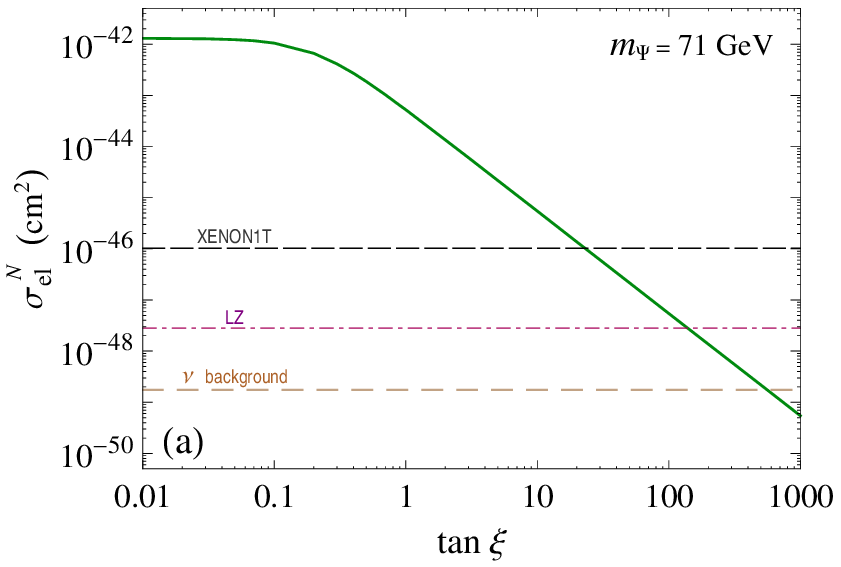}
\includegraphics[width=83mm]{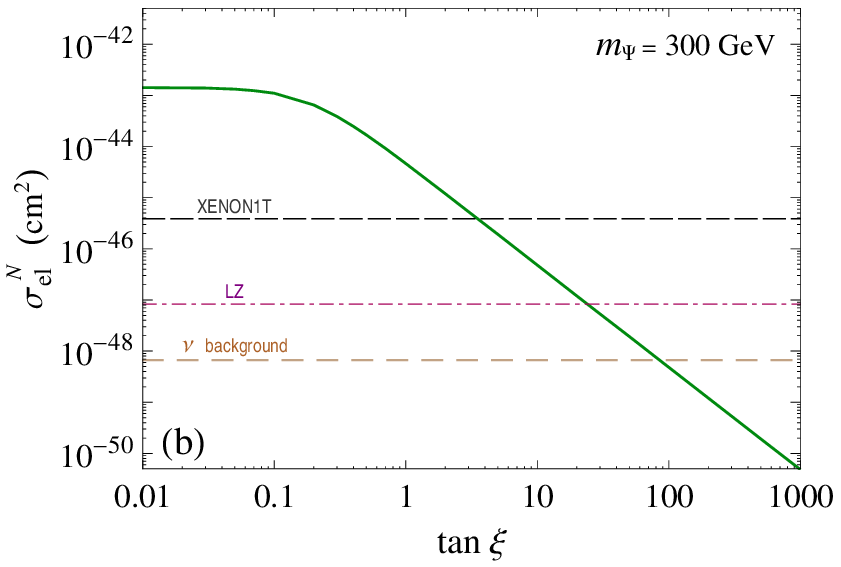}\vspace{-7pt}
\caption{The $\Psi$-nucleon cross-sections, $\sigma_{\rm el}^{N}$, (green curves)
versus \,$\tan\xi=\lambda_{\tt P}/\lambda_{\tt S}$\, for (a)\,\,$m_\Psi=71$\,\,GeV\, and
(b)\,\,$m_\Psi=300$\,\,GeV,\, compared to the measured upper-limit from XENON1T~\cite{x1t} as
well as the sensitivity projection~\cite{Cushman:2013zza} of LZ~\cite{lz} and
the neutrino background floor~\cite{nubg}.} \label{sigmaeltanxi}
\end{figure}

To provide some more insight into the dependence of $\sigma_{\rm el}^N$ on \,$\tan\xi$,\,
we give examples in Fig.\,\,\ref{sigmaeltanxi} for (a)\,\,$m_\Psi=71$\,\,GeV,\,
approximately corresponding to the peaks of the green solid curves for $\sigma_{\rm el}^N$
in Fig.\,\,\ref{sm+fdm-plots}, and
(b)\,\,$m_\Psi=300$\,\,GeV,\, which lies in the flat sections of the green curves.
From Figs.\,\,\ref{sm+fdm-plots} and\,\,\ref{sigmaeltanxi}, we conclude that for
\,$m_\Psi>300$\,GeV\, and \,$\tan\xi>100$\, the predicted $\sigma_{\rm el}^N$ is under
the neutrino-background floor.

\section{Comparison with Higgs-portal spin-1/2 DM\label{comparison}}

It is instructive to look at the differences and similarities between the previous scenario
and one in which the SM is slightly expanded with the inclusion of a spin-1/2 Dirac fermion
$\psi$ which is a singlet under the SM gauge group and serves as the WIMP DM candidate.
It is stable due to the $Z_2$ symmetry under which only $\psi$ is odd.
The DM Lagrangian with leading-order Higgs-portal couplings is then~\cite{deSimone:2014pda,
Beniwal:2015sdl,Escudero:2016gzx,Arcadi:2017kky,Kamenik:2011vy,Kamenik:2012hn,
LopezHonorez:2012kv}\footnote{A number of possibilities for the UV completion of this model
have been proposed in \cite{LopezHonorez:2012kv,Ghorbani:2014qpa,Tsai:2013bt}.
In a more complete theory, there may additionally be a spinless mediator which has renormalizable
pseudoscalar couplings not only to the DM, but also to SM quarks~\cite{Tsai:2013bt,pseudoscalar},
the latter of which induce spin-dependent DM-nucleon interactions.
The DM in this case is again elusive with regards to direct
detection~\cite{Tsai:2013bt,pseudoscalar}, unless the mediator is sufficiently
light~\cite{pseudoscalar} and/or also has a relatively sizable scalar coupling to the DM.}
\begin{eqnarray} \label{L-psi0}
{\cal L}_\psi^{} \,=\, \overline{\psi} \big( i\slash{\!\!\!\partial}
- \mu_\psi^{} \big) \psi - \Bigg( \frac{\overline{\psi}\psi}{\bar\Lambda}
+ \frac{i\overline{\psi}\gamma_5^{}\psi}{\bar\Lambda_5} \Bigg)
\textsf{\slshape H}^{\,\dagger\!}\textsf{\slshape H} \,,
\end{eqnarray}
where $\mu_\psi$, $\bar\Lambda$, and $\bar\Lambda_5$ are real constants.
The case in which the $\bar\Lambda_5$ term is absent has already been treated very recently
in Ref.\,\cite{Chang:2017gla} under the assumption of $CP$ invariance.
After electroweak symmetry breaking and the transformation of $\psi$ to the mass
eigenstate, we have
\begin{eqnarray} \label{L-psi}
{\cal L}_\psi^{}  \,\supset\, -m_{\psi\,}^{}\overline{\psi}\psi
\,-\, \overline{\psi} \big( \kappa_{\tt S}^{}
+ i\gamma_{5\,}^{}\kappa_{\tt P}^{} \big) \psi\, \bigg(h+\frac{h^2}{2v}\bigg) \,,
\end{eqnarray}
where
\begin{eqnarray}
m_\psi^{} &=& \sqrt{\bigg(\mu_\psi^{}+\frac{v^2}{2\bar\Lambda}\bigg)^{\!\!2}
+ \frac{v^4}{4\bar\Lambda_5^2}} ~, ~~~~ ~~~
\kappa_{\tt S}^{} \,=\,
\frac{v\cos\bar\theta}{\bar\Lambda}+\frac{v\sin\bar\theta}{\bar\Lambda_5} \,, ~~~~ ~~~
\kappa_{\tt P}^{} \,=\,
\frac{v\cos\bar\theta}{\bar\Lambda_5} - \frac{v\sin\bar\theta}{\bar\Lambda} \,,
\nonumber \\
\cos\bar\theta &=&
\frac{\mu_\psi^{}}{m_\psi^{}}+\frac{v^2}{2\bar\Lambda m_\psi^{}} \,, ~~~~ ~~~
\sin\bar\theta \,=\, \frac{v^2}{2\bar\Lambda_5^{}m_\psi^{}} \,.
\end{eqnarray}

We can then derive the cross section $\sigma_{\rm ann}^{}$ of the DM annihilation given by
\begin{eqnarray} && \hspace{-7em}
\sigma_{\rm ann}^{} \,=\, \sigma\big(\bar\psi\psi\to h^*\to X_{\textsc{sm}}\big) \,+\,
\sigma\big(\bar\psi\psi\to hh\big) \,, \vphantom{|_{\int_\int^{}}}
\nonumber \\
\sigma\big(\bar\psi\psi\to h^*\to X_{\textsc{sm}}\big) &=&
\frac{\big( \beta_{\psi\,}^{}\kappa_{\tt S}^2+\beta_\psi^{-1}\kappa_{\tt P}^2 \big)
\sqrt s~\raisebox{3pt}{\footnotesize$\displaystyle\sum_i$}\,
\Gamma\big(\tilde h\to X_{i,\textsc{sm}}\big)}
{2\big[ \big(s-m_h^2\big)\raisebox{1pt}{$^2$} + \Gamma_h^2m_h^2 \big]} \,,
\nonumber \\
\sigma\big(\bar\psi\psi\to hh\big) &=&
\frac{\beta_h^{}\big(\beta_\psi^{}\kappa_{\tt S}^2+\beta_\psi^{-1}\kappa_{\tt P}^2\big)}
{64\pi_{\,}v^2} \big({\cal R}_{hh}^2+{\cal I}_{hh}^2\big) \,,
\end{eqnarray}
the rate of the invisible decay \,$h\to\bar\psi\psi$\,
\begin{eqnarray}
\Gamma\big(h\to\bar\psi\psi\big) \,=\, \frac{m_h^{}}{8\pi} \Big[
\kappa_{\tt S}^2\, \big(1-4{\texttt{\slshape R}}_\psi^2\big)^{3/2}
+ \kappa_{\tt P}^2 \big(1-4{\texttt{\slshape R}}_\psi^2\big)^{1/2} \Big] \,,
\end{eqnarray}
and the cross section of $\psi$-nucleon elastic scattering
\begin{eqnarray}
\sigma_{\rm el}^N \,=\,\frac{g_{NNh\,}^2m_{\psi\,}^2 m_N^2}
{\pi_{\,}\big(m_\psi^{}+m_N^{}\big)\raisebox{0.7pt}{$^2$}m_h^4}
\Bigg[ \kappa_{\tt S}^2 + \frac{\kappa_{\tt P}^2\,m_{N\,}^2v_{\psi,\rm lab}^2}
{2\big(m_\psi^{}+m_N^{}\big)\raisebox{0.7pt}{$^2$}} \Bigg] \,,
\end{eqnarray}
where now \,$\Gamma_h^{}=\Gamma_h^{\textsc{sm}}+\Gamma\big(h\to\bar\psi\psi\big)$\,
and \,$v_{\psi,\rm lab}=10^{-3}$.\,
Like in the last section, one notices here that in the nonrelativistic limit
the $\lambda_{\tt S}$ ($\lambda_{\tt P}$) part of the annihilation rate ($\psi$-nucleon
cross-section) is substantially suppressed compared to its $\lambda_{\tt P}$
($\lambda_{\tt S}$) part.
This feature of the Higgs-portal spin-1/2 DM is well known in
the literature~\cite{deSimone:2014pda,Beniwal:2015sdl,Escudero:2016gzx,Arcadi:2017kky,
LopezHonorez:2012kv}.

In Fig.\,\,\ref{sm+fdm-plots} we have also provided examples for this spin-1/2 DM with
\,$\kappa_{\tt P}/\kappa_{\tt S}=\tan\xi=0,1,500$.\,
The blue curves (labeled 1/2) in the left panels represent the values of
\,$\lambda_{\psi h}=\big(\kappa_{\tt S}^2+\kappa_{\tt P}^2\big)\raisebox{0.7pt}{$^{1/2}$}$\,
consistent with the observed relic abundance and in the right panels the corresponding
predictions for the \mbox{$\psi$-nucleon} cross-section, $\sigma_{\rm el}^N$.
Analogously to their spin-3/2 counterparts, the LHC constraint on the Higgs invisible decay
implies that \,$\Gamma\big(h\to\bar\psi\psi\big)<0.16\,\Gamma_h$,\, and the EFT
limitation can be expressed as \,$\lambda_{\psi h}<v/m_\psi^{}$.\,
These are depicted in the left panels by the black dotted curves (labeled 1/2 as well) and
the magenta dashed curves, respectively.
The $\lambda_{\psi h}$ values in the $m_\psi$ ranges that fulfill these two requirements
translate into the solid parts of the blue curves for $\sigma_{\rm el}^N$ in the right panels.
From our numerical explorations, we learn that for \,$m_\psi>50$\,GeV\, and
\,$\kappa_{\tt P}^{}>20\,\kappa_{\tt S}^{}$\, the predicted $\sigma_{\rm el}^N$ is below all
the current bounds from direct searches and may even evade future ones, but the LHC and EFT
constraints reduce the allowed mass zone to
\,54\,\,GeV\,$\mbox{\footnotesize$\lesssim$}\,m_\psi\,
\mbox{\footnotesize$\lesssim$}\;$3.2\,\,TeV,\,
as may also be inferred from Fig.\,\,\ref{sm+fdm-plots}.

From the green and blue (solid) curves in Fig.\,\,\ref{sm+fdm-plots}, we can make some
comparison between these two models.
It is obvious that they resemble each other phenomenologically, but the viable
parameter space of the spin-3/2 DM is somewhat smaller.
It follows that new experimental limits on one of the models will likely apply to
the other in a similar manner.

Because of their similarities, to differentiate the two scenarios would require a high
degree of experimental precision.
Particularly, if confirmed positive signals in direct searches identify the mass of the DM to
be slightly below $m_h/2$ and its implied coupling to the Higgs is determined to be, for instance,
between the black dashed curves in Fig.\,\,\ref{sm+fdm-plots}(c), improved measurements
on the Higgs invisible decay can check the Higgs-portal hypothesis that the DM has spin 1/2.
This would likely be achievable at the High Luminosity LHC, which is expected to probe
the branching fraction of the invisible Higgs decay down to about
6\% at 95\% CL~\cite{Okawa:2013hda}.
For a significantly smaller DM-Higgs coupling, one would need the International Linear
Collider which could be sensitive to an invisible Higgs branching-fraction
as low as 0.4\% at 95\% CL~\cite{Asner:2013psa}.
If the DM mass is bigger than $m_h/2$ and away from the Higgs-pole region, it would likely be
more difficult for a collider probe to differentiate the two scenarios because the cross
section would be more suppressed by the Higgs propagator.
We note that a polarization measurement would not help much to discriminate them because
in each of them the DM interacts in pair with a Higgs via $s$-wave couplings at leading order.
In more complete theories, where the DM interactions could be more complicated and might
involve additional particles, we would expect that polarization studies and other methods
could be useful to distinguish the two DM candidates.

\begin{figure}[h]
\includegraphics[width=57mm]{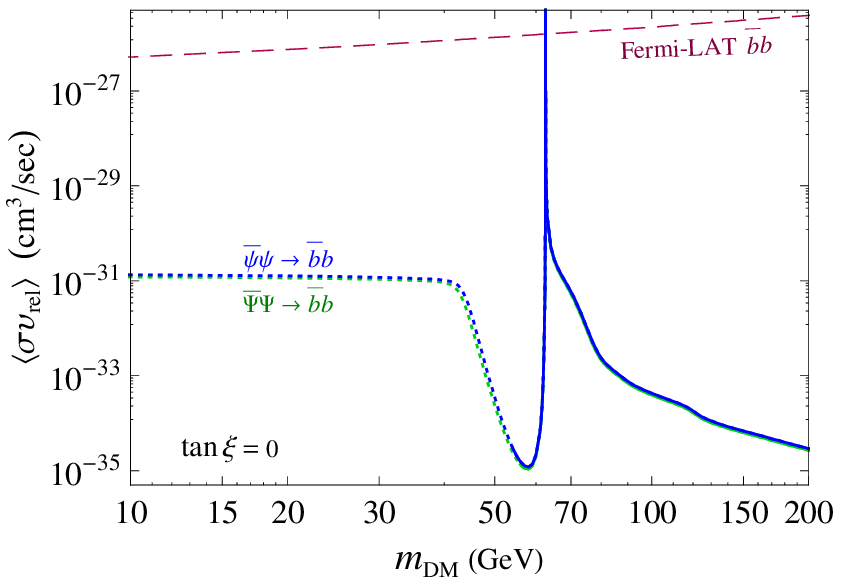}\!\!
\includegraphics[width=57mm]{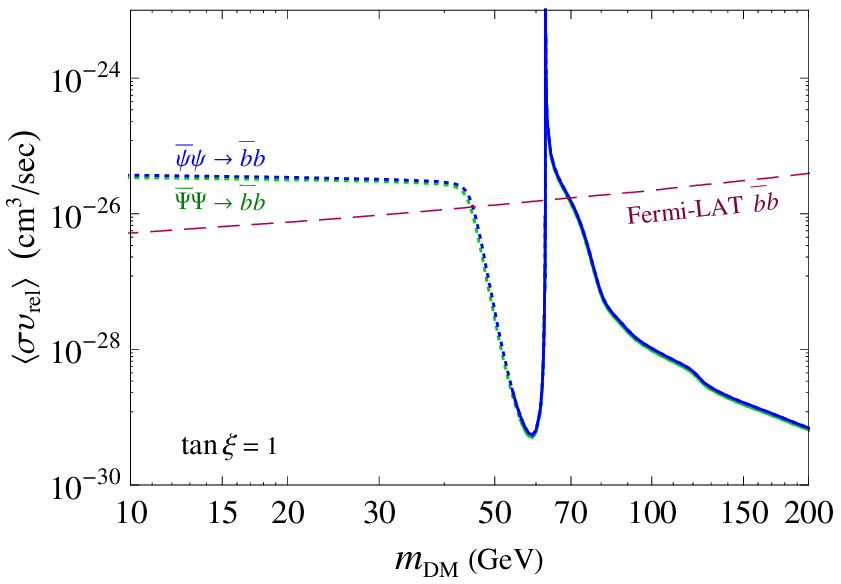}\!\!
\includegraphics[width=57mm]{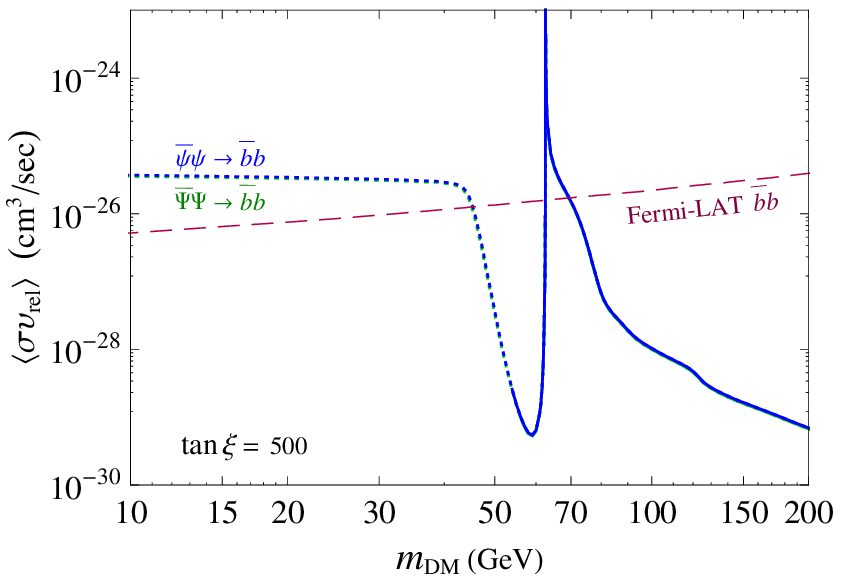}\vspace{-7pt}
\caption{The thermally averaged annihilation rates of \,$\bar\Psi\Psi\to\bar bb$\, and
\,$\bar\psi\psi\to\bar bb$\, compared to the corresponding Fermi-LAT bound~\cite{fermilat}.}
\label{indirect} \end{figure}

Since there is ample parameter space in these models that can escape upcoming direct
searches, even after the LHC and EFT constraints are imposed, it is of interest to
consider potential bounds from indirect detection experiments.
They may offer further checks, for the presence of the DM-Higgs pseudoscalar couplings,
whose effects on the annihilation rates do not suffer from $v_{\rm rel}$ suppression,
renders the DM potentially more observable in indirect quests.
At present we find a\,\,complementary restriction only from the results of searches for DM
annihilation signals from the Milky Way dwarf spheroidal galaxies with 6 years of Fermi
Large Area Telescope (Fermi-LAT) data~\cite{fermilat}.
The strongest limit occurs in the $\bar bb$ channel, but only for masses between 60 and
70\,\,GeV\, and preferably \,$\tan\xi\mbox{\footnotesize$\,\gtrsim\,$}1$,\,
as displayed in Fig.\,\,\ref{indirect} for the two models.
We expect that improved data in the future from Fermi-LAT~\cite{Charles:2016pgz} and other
efforts, such as the Cherenkov Telescope Array~\cite{Doro:2012xx}, will help probe these
models more stringently.

\section{Conclusions\label{conclusion}}

We have explored a simple WIMP DM scenario in which the DM candidate has spin 3/2 and
Higgs-portal interactions induced by effective dimension-five scalar and pseudoscalar
operators involving the standard Higgs doublet.
This kind of DM has not received as much attention as its spin-1/2 or bosonic counterparts
recently, but is a viable and interesting alternative, especially in light of the ongoing
Higgs measurements at the LHC and the continuing quests for DM with greatly improving
sensitivity.
As our examples demonstrate, the inclusion of the pseudoscalar operator in addition to
the scalar one is crucial for the model to avoid the existing strong restrictions from
direct detection experiments and possibly to evade future ones as well.
We also implemented the restraints from Higgs data and EFT considerations, which decrease
the allowed DM-mass region to between 58\,GeV and 2.3\,TeV.
We obtained an extra constraint from indirect searches by Fermi-LAT which disfavors
masses between 60 and 70\,\,GeV if the pseudoscalar coupling is not small relative
to the scalar one.
Finally, with explicit illustrations we showed that this model is similar in its important aspects
to its spin-1/2 counterpart, but has viable parameter space that is slightly more reduced.
Therefore, the combination of future data from LHC measurements and DM (in)direct searches
can be expected to test these two models more comprehensively.

As a last note, we mention that after this paper was submitted for publication
the PandaX-II Collaboration announced their newest upper-limit on the spin-independent
WIMP-nucleon cross-section~\cite{pandaxnew}.
For a WIMP mass exceeding 100 GeV this limit is the most stringent to date, but in the Higgs-pole
region (near $m_h/2$) it is almost identical to that from XENON1T~\cite{x1t}.
Hence the new PandaX-II results~\cite{pandaxnew} have not yet changed the viability of
the most minimal Higgs-portal fermionic (spin 3/2 or 1/2) WIMP DM, which we treated above.

\acknowledgments

This work was supported in part by the Ministry of Education (MOE) Academic Excellence Program
(Grant No. 105R891505) and National Center for Theoretical Sciences (NCTS) of the Republic of
China (ROC).
X.G.H was also supported in part by the Ministry of Science and Technology (MOST) of ROC
(Grant No. MOST104-2112-M-002-015-MY3) and in part by and in part by the National Science
Foundation of China (NSFC) (Grant Nos. 11175115, 11575111, and 11735010), Key Laboratory for
Particle Physics, Astrophysics and Cosmology, Ministry of Education, and Shanghai Key
Laboratory for Particle Physics and Cosmology (SKLPPC) (Grant No. 15DZ2272100) of the People's
Republic of China (PRC).


\begin{thebibliography}{0}

\bibitem{pdg}
  C.~Patrignani {\it et al.} [Particle Data Group],
  Chin.\ Phys.\ C {\bf 40}, no. 10, 100001 (2016).

\bibitem{Feng:2010gw}
  J.L.~Feng,
  Ann.\ Rev.\ Astron.\ Astrophys.\  {\bf 48}, 495 (2010)  [arXiv:1003.0904 [astro-ph.CO]].

\bibitem{Baek:2014jga}
  S.~Baek, P.~Ko, and W.I.~Park,
  Phys.\ Rev.\ D {\bf 90}, no. 5, 055014 (2014)  [arXiv:1405.3530 [hep-ph]].

\bibitem{He:2016mls}
  X.G.~He and J.~Tandean,
  JHEP {\bf 1612}, 074 (2016)  [arXiv:1609.03551 [hep-ph]].

\bibitem{Busoni:2014sya}
G.~Busoni, A.~De Simone, J.~Gramling, E.~Morgante, and A.~Riotto,
  JCAP {\bf 1406}, 060 (2014)  [arXiv:1402.1275 [hep-ph]].

\bibitem{deSimone:2014pda}
  A.~De Simone, G.F.~Giudice, and A.~Strumia,
  JHEP {\bf 1406}, 081 (2014)  [arXiv:1402.6287 [hep-ph]];
  M.A.~Fedderke, J.Y.~Chen, E.W.~Kolb, and L.T.~Wang,
  JHEP {\bf 1408}, 122 (2014)  [arXiv:1404.2283 [hep-ph]].

\bibitem{Beniwal:2015sdl}
  A.~Beniwal, F.~Rajec, C.~Savage, P.~Scott, C.~Weniger, M.~White, and A.G.~Williams,
  Phys.\ Rev.\ D {\bf 93}, no. 11, 115016 (2016)  [arXiv:1512.06458 [hep-ph]].

\bibitem{Escudero:2016gzx}
  M.~Escudero, A.~Berlin, D.~Hooper, and M.X.~Lin,
  JCAP {\bf 1612}, no. 12, 029 (2016)  [arXiv:1609.09079 [hep-ph]].

\bibitem{Arcadi:2017kky}
G.~Arcadi, M.~Dutra, P.~Ghosh, M.~Lindner, Y.~Mambrini, M.~Pierre, S.~Profumo, and F.S.~Queiroz,
  arXiv:1703.07364 [hep-ph].

\bibitem{Savvidy:2012qa}
  K.G.~Savvidy and J.D.~Vergados,
  Phys.\ Rev.\ D {\bf 87}, no. 7, 075013 (2013)  [arXiv:1211.3214 [hep-ph]].

\bibitem{gravitino} 
  L.~Roszkowski, S.~Trojanowski, K.~Turzynski, and K.~Jedamzik,
  JHEP {\bf 1303}, 013 (2013)  [arXiv:1212.5587 [hep-ph]];
  J.~Hasenkamp and M.W.~Winkler,
  Nucl.\ Phys.\ B {\bf 877}, 419 (2013)  [arXiv:1308.2678 [hep-ph]];
  K.~Benakli, Y.~Chen, E.~Dudas, and Y.~Mambrini,
  Phys.\ Rev.\ D {\bf 95}, no. 9, 095002 (2017)  [arXiv:1701.06574 [hep-ph]];
  E.~Dudas, Y.~Mambrini, and K.~Olive,
  Phys.\ Rev.\ Lett.\  {\bf 119}, no. 5, 051801 (2017)  [arXiv:1704.03008 [hep-ph]].

\bibitem{Dutta:2015ega}
  S.~Dutta, A.~Goyal, and S.~Kumar,
  JCAP {\bf 1602}, no. 02, 016 (2016)  [arXiv:1509.02105 [hep-ph]];
  M.O.~Khojali, A.~Goyal, M.~Kumar, and A.S.~Cornell,
  Eur.\ Phys.\ J.\ C {\bf 77}, no. 1, 25 (2017)  [arXiv:1608.08958 [hep-ph]].

\bibitem{Kamenik:2011vy}
  J.F.~Kamenik and C.~Smith,
  JHEP {\bf 1203}, 090 (2012)  [arXiv:1111.6402 [hep-ph]].

\bibitem{Kamenik:2012hn}
J.F.~Kamenik and C.~Smith,
  Phys.\ Rev.\ D {\bf 85}, 093017 (2012)  [arXiv:1201.4814 [hep-ph]].

\bibitem{Yu:2011by}
  Z.H.~Yu, J.M.~Zheng, X.J.~Bi, Z.~Li, D.X.~Yao, and H.~H.~Zhang,
  Nucl.\ Phys.\ B {\bf 860}, 115 (2012)  [arXiv:1112.6052 [hep-ph]];
  R.~Ding and Y.~Liao,
  JHEP {\bf 1204}, 054 (2012)  [arXiv:1201.0506 [hep-ph]];
  R.~Ding, Y.~Liao, J.Y.~Liu, and K.~Wang,
  JCAP {\bf 1305}, 028 (2013)  [arXiv:1302.4034 [hep-ph]].

\bibitem{Chang:2017gla}
  C.F.~Chang, X.G.~He, and J.~Tandean,
  JHEP {\bf 1704}, 107 (2017)  [arXiv:1702.02924 [hep-ph]].

\bibitem{lux} 
  D.S.~Akerib {\it et al.},
  Phys.\ Rev.\ Lett.\  {\bf 118}, no. 2, 021303 (2017)  [arXiv:1608.07648 [astro-ph.CO]].

\bibitem{pandax} 
  A.~Tan {\it et al.} [PandaX-II Collaboration],
Phys.\ Rev.\ Lett.\  {\bf 117}, no. 12, 121303 (2016)  [arXiv:1607.07400 [hep-ex]].

\bibitem{x1t} 
  E.~Aprile {\it et al.} [XENON Collaboration],
Phys.\ Rev.\ Lett.\  {\bf 119}, no. 18, 181301 (2017)  [arXiv:1705.06655 [astro-ph.CO]].

\bibitem{dmdd} 
  R.J.~Gaitskell,
  Ann.\ Rev.\ Nucl.\ Part.\ Sci.\  {\bf 54}, 315 (2004);
  J.~Kopp, T.~Schwetz, and J.~Zupan,
  JCAP {\bf 1002}, 014 (2010)  [arXiv:0912.4264 [hep-ph]];
  P.~Panci,
  Adv.\ High Energy Phys.\  {\bf 2014}, 681312 (2014)  [arXiv:1402.1507 [hep-ph]];
  E.~Aprile {\it et al.} [XENON Collaboration],
  JCAP {\bf 1604}, no. 04, 027 (2016)  [arXiv:1512.07501 [physics.ins-det]].

\bibitem{Cushman:2013zza}
  P.~Cushman {\it et al.},
  arXiv:1310.8327 [hep-ex].

\bibitem{Aad:2015pla}
  G.~Aad {\it et al.} [ATLAS Collaboration],
  JHEP {\bf 1511}, 206 (2015)  [arXiv:1509.00672 [hep-ex]].
  V.~Khachatryan {\it et al.} [CMS Collaboration],
  JHEP {\bf 1702}, 135 (2017)  [arXiv:1610.09218 [hep-ex]].

\bibitem{atlas+cms}
The ATLAS and CMS Collaborations,
  JHEP {\bf 1608}, 045 (2016)  [arXiv:1606.02266 [hep-ex]].

\bibitem{rs} 
  W.~Rarita and J.~Schwinger,
  Phys.\ Rev.\  {\bf 60}, 61 (1941).

\bibitem{Moldauer:1956zz}
  P.A.~Moldauer and K.M.~Case,
  Phys.\ Rev.\  {\bf 102}, 279 (1956);
C. Fronsdal, Nuovo Cimento Suppl. 9, 416 (1958).

\bibitem{Christensen:2013aua}
  N.D.~Christensen {\it et al.},
  Eur.\ Phys.\ J.\ C {\bf 73}, no. 10, 2580 (2013)  [arXiv:1308.1668 [hep-ph]];

\bibitem{polsum}
Y. Takahashi and H. Umezawa,
Progr. Theoret. Phys. (Kyoto) {\bf 9}, 14 (1953);
  R.E.~Behrends and C.~Fronsdal,
  Phys.\ Rev.\  {\bf 106}, no. 2, 345 (1957).

\bibitem{lhc:mh}
  G.~Aad {\it et al.}  [ATLAS and CMS Collaborations],
  Phys.\ Rev.\ Lett.\  {\bf 114}, 191803 (2015)  [arXiv: 1503.07589 [hep-ex]].

\bibitem{lhctwiki}
  S.~Heinemeyer {\it et al.}  [LHC Higgs Cross Section Working Group Collaboration],
  arXiv:1307.1347 [hep-ph].
Online updates available at \\
https://twiki.cern.ch/twiki/bin/view/LHCPhysics/CERNYellowReportPageBR2014.

\bibitem{planck} 
  P.A.R.~Ade {\it et al.} [Planck Collaboration],
  Astron.\ Astrophys.\  {\bf 594}, A13 (2016)  [arXiv:1502.01589 [astro-ph.CO]].

\bibitem{dsg2} 
  C.E.~Aalseth {\it et al.},
  Adv.\ High Energy Phys.\  {\bf 2015}, 541362 (2015).

\bibitem{lz} 
  D.S.~Akerib {\it et al.} [LZ Collaboration],
  arXiv:1509.02910 [physics.ins-det].

\bibitem{nubg} 
  J. Billard, L. Strigari, and E. Figueroa-Feliciano,
  Phys.\ Rev.\ D {\bf 89}, no. 2, 023524 (2014)  [arXiv:1307.5458 [hep-ph]].

\bibitem{LopezHonorez:2012kv}
  L.~Lopez-Honorez, T.~Schwetz, and J.~Zupan,
  Phys.\ Lett.\ B {\bf 716}, 179 (2012)  [arXiv:1203.2064 [hep-ph]].

\bibitem{Ghorbani:2014qpa}
  K.~Ghorbani,
  JCAP {\bf 1501}, 015 (2015)  [arXiv:1408.4929 [hep-ph]];
  Y.G.~Kim, K.Y.~Lee, C.B.~Park, and S.~Shin,
  Phys.\ Rev.\ D {\bf 93}, no. 7, 075023 (2016)  [arXiv:1601.05089 [hep-ph]].

\bibitem{Tsai:2013bt}
  H.C.~Tsai and K.C.~Yang,
  Phys.\ Rev.\ D {\bf 87}, no. 11, 115016 (2013)  [arXiv:1301.4186 [hep-ph]];
  S.~Baek, P.~Ko, and J.~Li,
  Phys.\ Rev.\ D {\bf 95}, no. 7, 075011 (2017)  [arXiv:1701.04131 [hep-ph]].

\bibitem{pseudoscalar} 
  C.~Boehm, M.J.~Dolan, C.~McCabe, M.~Spannowsky, and C.J.~Wallace,
  JCAP {\bf 1405}, 009 (2014)  [arXiv:1401.6458 [hep-ph]];
  C.~Arina, E.~Del Nobile, and P.~Panci,
  Phys.\ Rev.\ Lett.\  {\bf 114}, 011301 (2015)  [arXiv:1406.5542 [hep-ph]].

\bibitem{Okawa:2013hda}
  H.~Okawa, J.~Kunkle, and E.~Lipeles,
  arXiv:1309.7925 [hep-ex].

\bibitem{Asner:2013psa}
  D.M.~Asner {\it et al.},
  arXiv:1310.0763 [hep-ph].

\bibitem{fermilat} 
  M.~Ackermann {\it et al.} [Fermi-LAT Collaboration],
  Phys.\ Rev.\ Lett.\  {\bf 115}, no. 23, 231301 (2015)  [arXiv:1503.02641 [astro-ph.HE]].

\bibitem{Charles:2016pgz}
  E.~Charles {\it et al.} [Fermi-LAT Collaboration],
  Phys.\ Rept.\  {\bf 636}, 1 (2016)  [arXiv:1605.02016 [astro-ph.HE]].

\bibitem{Doro:2012xx}
  M.~Doro {\it et al.} [CTA Consortium],
  Astropart.\ Phys.\  {\bf 43}, 189 (2013)  [arXiv:1208.5356 [astro-ph.IM]].

\bibitem{pandaxnew}
X. Cui {\it et al.} [PandaX-II Collaboration],
  Phys.\ Rev.\ Lett.\  {\bf 119}, no. 18, 181302 (2017)  [arXiv:1708.06917 [astro-ph.CO]].

\end{thebibliography}
\end{document}